\documentclass[aps,pra,twocolumn,groupedaddress,amssymb,amsfonts,showpacs]{revtex4}

\usepackage{color}
\usepackage[dvips]{graphics,graphicx}
\usepackage{calc,ifthen}

\newlength{\elimdepthdim}
\newlength{\elimheightdim}
\newlength{\elimwidthdim}

\newlength{\strutdepthdim}
\newlength{\strutheightdim}
\newlength{\strutwidthdim}

\def\one{{\mathchoice {\rm 1\mskip-4mu l} {\rm 1\mskip-4mu l} {\rm
1\mskip-4.5mu l} {\rm 1\mskip-5mu l}}}

\newcommand{\ket}[1]{\qvbar{#1}\qrangle}

\newcommand{\cE}{{\cal E}}

\newcommand{\cS}{{\cal S}}

\newcounter{herefignum}

\newcommand{\shortqph}[1]{}

\providecommand{\ignore}[1]{}

\def\openone{\leavevmode\hbox{\small1\kern-3.8pt\normalsize1}}
\def\RR{{\rm I\kern-.2emR}}

\def\openone{\leavevmode\hbox{\small1\kern-3.8pt\normalsize1}}
\def\RR{{\rm I\kern-.2emR}}

\def\ce{{\cal E}}

\def\cs{{\cal S}}

\providecommand{\ignore}[1]{}

\renewcommand{\ket}[1]{| #1 \rangle}

\newcommand{\bitem}{\begin{itemize}}
\newcommand{\eitem}{\end{itemize}}
\newcommand{\benum}{\begin{enumerate}}
\newcommand{\eenum}{\end{enumerate}}
\newcommand{\beq}{\begin{equation}}
\newcommand{\eeq}{\end{equation}}
\newcommand{\beqa}{\begin{eqnarray}}
\newcommand{\eeqa}{\end{eqnarray}}
\newtheorem{definition}{Definition}

\newtheorem{proposition}{Proposition}

\newcommand{\bproof}{\begin{proof}}
\newcommand{\eproof}{\end{proof}}
\newcommand{\bprop}{\begin{proposition}}

\newcommand{\bdef}{\begin{definition}}

\begin{document}


\title{No-signalling-based version of Zurek's derivation of
quantum probabilities:  A note on ``Environment-assisted invariance, 
entanglement, and probabilities in quantum physics''}


\author{Howard Barnum}
\affiliation{$^1$CCS-3, Mail Stop B256, Los Alamos National Laboratory,
Los Alamos, NM 87545 {\tt barnum@lanl.gov}}



\date{\today}

\begin{abstract}
Zurek  has derived the quantum probabilities for
Schmidt basis states of bipartite quantum systems in pure joint
states, from the assumption that they should be not be affected by one
party's action if the action can be undone by the other party
(``envariance of probability'') and an auxiliary assumption.  We argue
that a natural generalization of the
auxiliary assumption is actually strong enough to yield the
Born rule itself, but that Zurek's argument and protocol can be
adapted to do without this assumption, at the cost of using 
envariance of probability in both directions.  We consider alternative
motivations for envariance, one based on the no-signalling constraint that
actions on one subsystem of a quantum system not allow signalling
to another subsystem entirely distinct from the first, and another which
is perhaps strongest in the context of a relative-state interpretation of
quantum mechanics.  In part because of this, we argue that 
the relative appeal of our version and the original 
version of Zurek's argument depends in part upon
whether one interprets the quantum formalism
in terms of relative states or definite measurement outcomes. 
\end{abstract}



\maketitle

In a recent paper \cite{Zurek2003a}, Wojciech Zurek provides a
derivation of the Born rule for quantum probabilities from other
aspects of quantum mechanics, and an auxiliary
assumptions.  Central to his argument is the possibility of
``envariance'': the existence of states of composite systems $SE$ and
nontrivial unitaries acting only one part of the system (say $S$),
such that the unitary's effect can be undone by acting on the other
part.  Zurek uses the assumption (which he also sometimes calls
``envariance,'' and which we will always try to qualify as
``envariance of probabilities''): that probabilities ascribed to
states of $S$ should be unaffected by a transformation $u_S$ on $S$
when the composite system is in a state envariant under $u_S$.  In
concert with a ``pedantic'' auxiliary assumption about probabilities,
this enables him to derive the usual quantum probability rule.  One
purpose of our comment is to point out that if both envariance and
no-signalling are formulated symmetrically with respect to 
interchange of subsystems $S$ and $E$, then (a) the auxiliary assumption
is actually rather strong, and a natural strengthening of it 
is strong enough to give the Born rule
directly, but (b) the protocol Zurek gives in his argument supports a
closely related argument that uses {\em only} envariance of
probabilities.  The envariance-based arguments establish quantum
probabilities only for subsystems of composite systems in entangled
states, but, especially within a relative-state viewpoint such as
Zurek's, this may well suffice for using quantum theory to make
predictions.   As we understand Zurek's argument,
it uses both assumptions, but each in only one direction, and we
also explore the possibility that, especially from the relative 
state point of view, each of the assumptions might be justified
only in the direction Zurek uses it.  From this point of view, 
our work clarifies both the formal structure of Zurek's argument
and its close relation to the relative state interpretation, and
explores similar arguments exploiting 
``envariance,'' and their appeal given alternative interpretations.
We view the motivation for the symmetric version of 
envariance of probabilities as somewhat different
than Zurek's primary motivation: we view it as a ``no-signalling'' assumption.  
Of course, Zurek alludes to this motivation when he notes that no 
signalling from one subsystem to another (aka ``causality,'' especially when the
systems are spacelike separated) implies envariance of
probabilities. He views causality as a ``more costly assumption'' because
it explicitly assumes that instantaneous signalling is impossible for
{\em all} states and local transformations, not just envariant combinations
of them.  The version of Zurek's argument we give below uses envariance
in both directions;  we envision  using no-signalling as a motivation
for the envariance-of-probability assumptions used in the argument, and
strictly speaking do not require the full no-signalling assumption which,
as Zurek has pointed out, is in the above sense stronger.
We note, however, that if no-signalling is restricted to apply
not necessarily between {\em arbitrary} 
distinct subsystems of a composite
system, but only between spacelike-separated ones, the assumption
becomes neither logically weaker nor stronger than 
(and if restricted to envariant state-transformation 
combinations,
logically weaker than) Zurek's envariance assumption.   This raises the question of
whether, when restricted to ``no faster-than-light signalling,'' the assumption still
suffices to derive probabilities for the outcomes of all quantum experiments.
The motivations for envariance will be discussed further below.

As a preliminary, we note that Zurek often works within the Everett
``relative state'' interpretation \cite{Everett57a, Everett57b}, in
which quantum measurement is viewed as a situation in which an
observer $O$, who is possibly in contact with an environment $E$ and
possibly some apparatus $A$, evolves jointly with a ``measured''
system $S$.  Measurements are viewed as evolutions that correlate some
set of mutually orthogonal subspaces of $S$ (which may be viewed as
the eigenspaces of an observable being measured) with states, or at
least with orthogonal eigenspaces, which we call ``pointer spaces,''
in which the observer is having distinct experiences (e.g. seeing a
detector flash or fail to flash).  (This observer eigenspace
decomposition is likely also singled out by its tendency to
``decohere,'' i.e. become corelated with the environment.)  On this
interpretation there is no physical ``collapse'' associated with
measurement, but one may ask for probabilities assigned to the
different orthogonal statevector components associated with the
decomposition, representing the likelihood that one will subjectively
experience one's consciousness traveling along one of the branches
picked out by this decomposition (cf. \cite{Barnum90a}).  In our
opinion, the version of Zurek's argument we give below does not depend
crucially on whether measurement is interpreted in this way, or as
involving ``collapse,'' or in some other way (for example as involving
``collapse'' of our knowledge, say, in a process similar to Bayesian
updating \cite{Caves2002a}).  However, the choice of interpretation
might profoundly affect the motivation for making the assumptions used
in the argument.  The issue in the formal part of the argument is what
probabilities can consistently be attributed to orthogonal subspaces,
and it arises on any interpretation, though it is perhaps of more
concern on the relative state interpretation, within which there is
often felt a need to ``derive'' probabilities from the
nonprobabilistic statements about the nature of the state, as an
objective property of the universe, that constitute the core of the
theory; other approaches are often more comfortable with taking
explicit specification of probabilities as part of the theory.  From
the point of view of (modestly) formal arguments, the distinction
between this relative state approach and the others we consider lies
in the fact that with relative state, we seek to assign probabilities
to components of the statevector, whereas with the others, we seek a
rule assigning probabilities to mutually orthogonal subspaces in some
decomposition of the Hilbert space (given a statevector).  The
difference is subtle, and for our purposes may be essentially
obliterated by some arguments we give below (but these rely on
symmetric no-signaling).

Central to Zurek's argument is 
the peculiar nature of composite quantum systems, notably the
possibility of ``envariance.''  A state $\ket{\psi}$ of $\cs \ce$ 
is said to be {\em envariant} with respect to a unitary $u_\cs$ if there
exists a unitary $u_\ce$ such that 
\beqa
(\one \otimes u_\cE) (u_\cS \otimes \one) \ket{\psi} = \ket{\psi}\;.
\eeqa
That is, the effect of doing $u_S$ on a system when the joint state
of system and environment is $\ket{\psi}$ 
can be undone by some unitary acting only
on the environment.

Zurek's argument involves the assumption, ``envariance of
probabilities,'' that, given a state $\psi_{\cs\ce}$, any aspect of
the system that can be changed by a transformation under which the
state is envariant, should not affect the probabilities ascribed to
states of the system.  He notes that envariance assures {\em
causality}: if aspects of $\cs$ that can be influenced solely by
acting on $\ce$ affected probabilities in $\cs$, we could signal from
$\cs$ to $\ce$ by manipulating them in $\ce$.  Actually, we prefer to
call this ``no-signalling''; if the unitary $u_\ce$ and the
``measurement'' on $\cs$ whose probabilities it affects can be
performed fast enough, compared to the spacelike separation of $\cs$
and $\ce$, such signalling would violate causality, which is certainly
one of the main reasons for ruling it out.  Perhaps, however, there is
a stronger argument for no $S$-to-$E$ signalling in a relative state
interpretation.  On such an interpretation, once macroscopic aspects
of $E$ have been correlated with $S$ (the system has been ``measured''
by an observer who is part of $E$), the ability to affect
probabilities of components of the state in subspaces corresponding to
those distinct macroscopic aspects of $E$, by manipulating $S$,
jeopardizes the interpretation of these numbers as ``probabilities''
at all.  (At a minimum, it would require, to have probabilities for
the results of a measurement on $S$, that we know the entire
subsequent dynamics of $S$, and consistently ascribe the same
probabilities to all evolved decompositions after definite results are
observed in $E$-- something that seems very difficult to do---though
this is not a proof--- in any other way than by making probabilities a
function of the square modulus of the wavefunction component.)  One
might very well object that probabilities are not ``objective''
aspects of the situation, that they represent beliefs about how likely
things are to happen.  More precisely (within a generally subjectivist
approach to probability in its aspect as something to be {\em used} in
science and everyday life \cite{Savage72a, Fishburn70a, Kreps88a}, an
approach to which I am rather partial), one might say it represents
constraints imposed by rationality (and certain technical assumptions)
on the bets or decisions one might condition on outcomes of a quantum
or measurement (cf.  e.g. \cite{Pitowsky2003a}).  There may or may not
be difficulties with extending this subjectivist approach to the
relative-state interpretation.  The potential difficulty is that since
the ``outcomes'' are purely perspectival, from the vantage point of
the decision-maker before the branching of consciousness, the {\em
objective} situation after the decision, i.e. the full entangled state
vector, is the ``real'' outcome.  The goal must then be, \'a la
Deutsch \cite{Deutsch99a}, to attempt to show that a rational
decision-maker should behave ``as if'' the branches had probabilities
given by the Born rule.  From either of these points of view, however,
the argument that we are manipulating probabilities of macroscopically
distinct subspaces of $E$ by manipulating $S$ and it is is manifestly
undesirable that this affect the probabilities of outcomes, if it
retains any validity, has to be glossed as something like: a rational
choice between different experiments on $S$ should be unaffected by
how $S$ will be treated after the observer (as part of $E$)
``branches'' into macroscopically different conscious states.  Indeed,
in many cases when apparatus (possibly rapidly decohered by other
parts of the environment) mediates between observer and system, one
might also find it reasonable to assume that nothing done to $S$ after
it has become thus entangled with ``pointer subspaces'' in the
apparatus and environment, should affect the decision between
experiments.  (All this requires assumptions, typical in classical
decision theory as well and capable of being formalized, that the
system itself is not something the observer particularly cares
about---though this formalization is not a trivial matter, since it
requires distinguishing between ``not caring about manipulations on
the system because they don't affect things we care about'', and not
caring about them {\em per se}, i.e. assuming we know nothing about
how they affect other things.)  This sort of assumption starts to seem
rather closely related to some of the invariance-of-decision
assumptions made, for example, by Wallace \cite{Wallace2002a,
Wallace2003a} and on Wallace's interpretation of him, by Deutsch.  In
what follows, we use the language of ``probabilities for outcomes'' or
``probabilities for components of the state vector.''  It is true that
in the latter case, most naturally associated with the relative state
view, the assumption that rational decision-makers will behave {\em as
if} they had probabilities may be a substantive one, requiring
arguments of the sort put forth by Deutsch and Wallace.  (I will make
the side comment that from a decsion-theoretic point of view of
applied probability, it could be reasonable to maintain that to behave
as if one has probabilities is to have them.)

For those arguments that do not require interpretation as ``system'' and
``environment,'' we introduce Stan and Emma as counterparts of the more
usual Alice and Bob.  Stan and Emma may be thought of as characters who
operate on the systems $S$ and $E$.  
The first step of Zurek's argument is to point out that if the state is 
\beqa \label{Schmidt}
\ket{\psi^{SE}} = \sum_{k = 1}^N \alpha_k \ket{\sigma_k}
\ket{\varepsilon_k}\;,
\eeqa
where $\ket{\sigma_k}$ are some orthonormal basis for $S$, 
and $\ket{\varepsilon_k}$ one for $E$ (for some choice of such bases,
any pure state has this form), then ``an envariant description of 
the system (i.e. $S$) must ignore phases of the coefficients''
$\alpha_k$.  ``Such a description must be based on a set of 
pairs $\{|\alpha_k|, \ket{\sigma_k}\}$.''  That is, when the state has
this form---and any pure state does---probabilities involving the
system $S$ must be based only on the absolute values of coefficients,
and the basis elements they are attached to---exactly the information
contained in the reduced density matrix.
We consider this step to establish, from no-signalling (aka envariance
of probabilities), that when the joint state of $SE$ is pure, the 
probabilities of orthogonal subspaces in 
{\em any}  subspace decomposition of $S$---are independent of the
phases in the Schmidt expansion of $\ket{\psi}$.  Note that this
may be more than Zurek is concerned with:  he is concerned primarily
with the probabilities of the decomposition of $S$ into the states
$\sigma_k$, or $E$ into the states $\varepsilon_k$, for if the joint
state $SE$ is viewed as the outcome of a measurement
``in the Schmidt basis'' on $S$, by an environment $E$ that includes
the observer, whose ``definite measurement results'' line up with 
the Schmidt basis for $E$, ascribing probabilities to these suffices
for ascribing probabilities to ``definite measurement results'' 
(subjectively experienced, of course) in the relative state interpretation;
any measurement {\em gives rise to} such a situation.

Note that we have not yet established that, for a given state,
the probabilities of 
components in 
subspaces are {\em independent} of the subspace decomposition in
which they occur, an assumption similar to that made in Gleason's
theorem, and which might allow us to use Gleason's theorem as 
part of an argument for quantum probabilities.  Of course, a potential
virtue of the argument from envariance is precisely that it does
not make any such assumption to begin with.

Another possible assumption
is that in a decomposition such as (\ref{Schmidt}),
the probabilities of obtaining $k$ when measuring in the basis 
$\ket{\sigma_k}$ on $S$, and obtaining it when measuring in
the basis $\ket{\varepsilon_k}$ on $E$, are the same.
We call this the {\em Perfect Correlation Principle} (PCP).
Without a noncontextuality assumption, however, this is not immediately
evident, and needs to be established. 
The reason it seems plausible is that both correspond to the same component, 
$\ket{\sigma_k} \ket{\varepsilon_k}$.
 In other situations, the following
may be used.
Clearly, one typical way in 
which this measurement outcome occurs is when each party measures
some local observable of which it spans an eigenspace.  Whether 
or not Stan measures anything should be immaterial to 
Emma's probability, by no-signalling.  Hence the probability of 
Emma's getting $\ket{\psi_k}$ should not be affected by the context 
in which $\ket{\varepsilon_k}$ is measured on $E$, or even
whether it is measured at all.  Incidentally, this establishes
noncontextuality of probabilities {\em of Schmidt basis 
states} in local measurements, since
if the local context affected the probability, that changed 
probability would be detectable by the other party.  

(Possibly
some more elaborate variant of the above argument,
probably requiring no-signalling via arbitrary states, not
just Schmidt basis states, but
perhaps only in one direction, 
could establish noncontextuality of
arbitrary measurements, at least when the joint state is pure;
this could be worth looking into as an alternative derivation
of probabilities.)

The next step of Zurek's argument, unlike the first one, does
not extend to probabilities of arbitrary local measurements, 
but concerns ``the relation between the coefficients $\alpha_k$
of the corresponding set of the candidate pointer states 
$\{\ket{\sigma_k}\}$ and their probabilities.''  

The key part of this step involves a situation where all, or some,
of the $|\alpha_k|$ are equal.  A simplified version of it, capturing
everything essential, involves the following protocol (in which all
states appearing are joint states of $SE$, with $S$, appearing
on the left, locally acted
on by ``Stan'' and $E$, to the right, locally acted on by ``Emma'').
\beqa
\ket{\psi_1} = \frac{1}{\sqrt{2}}(\ket{0}\ket{0} + \ket{1}\ket{1}\;.\\
\downarrow \text{~ Emma swaps~}0,1\;.\nonumber \\
\ket{\psi_2} = 
\frac{1}{\sqrt{2}}(\ket{0}\ket{1} + \ket{1}\ket{1}\;.\\
\downarrow \text{~ Stan swaps~}0,1\;.\nonumber \\
\ket{\psi_3} = 
\frac{1}{\sqrt{2}}(\ket{1}\ket{1} + \ket{0}\ket{0} 
\equiv \ket{\psi_1}\;.
\eeqa
Here ``swaps $0,1$'' means, of course, performs the unitary transformation
whose matrix in the basis $\ket{0},\ket{1}$ is $\sigma_x$.
Emma's swap can be undone by an Stan unitary;  in Zurek's terms, Emma's
states $\ket{0}$ and $\ket{1}$ are {\em envariantly swappable}.

For use in the discussion below, 
we define probabilities $p^X(i|\psi_j)$, 
$i \in \{0,1\}$,
$j \in \{1,2,3\}$, $X \in \{A,B\}$, to be $X$'s (i.e. Stan's or Emma's)
probability for state $\ket{i}$, when the joint state is $\ket{\psi_j}$.
Note that some may prefer to assign probabilities to states or subspaces
of the overall system;  the probabilities just mentioned can be 
so interpreted, by no-signalling arguments of the type associated with 
the Perfect Correlation Principle.  The no-signalling assumption required
may, however, be stronger than Zurek's in that it is not restricted to
transformations under which the state is envariant (but that may already
apply to the one used in the derivation of the PCP itself).  In 
this application (contrary to the one above) no-signalling must be applied 
to non-Schmidt-basis measurements.

The PCP itself establishes that 
\beqa
p^S(0|\psi_1) = p^E(0|\psi_1), p^S(1|\psi_1) = p^E(1|\psi_1)\; ,\nonumber \\
p^S(0|\psi_2) = p^E(1|\psi_2), p^S(1|\psi_2) = p^E(0|\psi_2)\;.
\eeqa
Thus, for the purpose of interpreting our arguments below in terms
perhaps more congenial to relative state-er's and to Zurek, one
can identify, for example, both  $p^2_A(0)$ and $p^2_B(1)$
with the probability of $\ket{0}\ket{1}$ in state $\ket{\psi_2}$.

At this point, Zurek invokes ``a rather general (and pedantic)
assumption about the measuring process: When the states are swapped,
the corresponding probabilities get relabeled $(i \leftrightarrow j$).
This leads us to conclude that the probabilities for any two envariantly
swappable $\ket{\sigma_k}$ are equal.''  

We interpret this assumption to be about the relation of the
probabilities $p_i, p_j$ of states $i$ and $j$ of $S$ before the swap,
to the probabilities $p'_i, p'_j$ of outcomes $i$ and $j$ of $S$ after
a unitary that swaps $i$ and $j$ is performed on $S$.  The assumption
is that $p'_i = p_j$ and $p'_j = p_i$.  In its narrowest version, the
assumption might be meant only to apply to envariant swaps like those in
the protocol above.  Zurek
has described the assumption to us as being essentially
invariance of the probabilities under redirection of the detector
output channels.  The idea would seem to be that $i$ and $j$ are just
``arbitrary labels'' anyway.  Another description of it---using the
language of measurement---might be be ``swapping states and then
measuring an observable for which these states span eigenspaces, has
the same probabilities (for eigenvalues of the observable) as just
measuring the observable conjugated by the swap operator.''  Phrased
thus, it is a special case of an assumption that others have used to
derive the Born probabilities---which one might formulate as
``evolving and then measuring is measuring the evolved observable''
(EATMIMTEO).  The others include \cite{Wallace2003a, Wallace2002a}, 
who argues this
was what was really meant in \cite{Deutsch99a}, and perhaps also
\cite{Saunders2002a}.  In \cite{Barnum2000a} we characterized the
special case as ``akin to Laplace's principle of insufficient reason''
for probabilities (that for $N$ mutually exclusive events between
which we can see no relevant differences, the probabilities of each
event should be taken to be $1/N$), and Zurek has also mentioned
Laplace in this connection.  However, it is
probably preferable to avoid the reformulation as a case of EATMIMTEO,
for the formalism of assigning measurement results to eigenspaces of
observables, rather than to states in a decomposition of a state
vector, carries with it an automatic assumption that the phase of a
state cannot affect its probability, a matter that Zurek is at pains
to establish via an envariance argument.  (Of course, since the
assumptions of \cite{Wallace2002a, Wallace2003a} are ultimately interpreted in
relative state terms, Wallace's motivation for this
special case---or perhaps for the general case---of EATMIMTEO might
turn out to be essentially envariance-based, as well.)

However it is justified, the assumption allows {\em immediate}
derivation of the equality of the probabilities $p^S(0|\psi_1)$ 
and $p^S(1|\psi_1)$ above.
Indeed, if it is not restricted to envariant situations (and there
seems to be no reason to so restrict it, although 
Zurek's argument can live with such a restriction) it also gives 
$p(\ket{\psi_i}|\ket{\psi}) = p(\ket{\psi_j}|\ket{\psi})$ 
for $i,j \in \{1,...,N \}$,  for 
{\em any} state $\ket{\psi} = \sum_{k=1}^N \alpha_k \ket{\psi_k}$ in which
$\alpha_l = \alpha_m$ for all $l,m \in \{1,...,N\}$.

The use contemplated for the PA in Zurek's argument is perhaps as follows.
Consider the protocol above.
By the pedantic assumption, 
$p^E(0|\psi_1) = p^E(1|\psi_2)$.  By envariance of probabilities, 
$p^E(1|\psi_2) = p^E(1|\psi_3) \equiv p^E(1|\psi_1)$.
This establishes $p^E(0|\psi_1) = p^E(1|\psi_2)$, as desired (so 
each is equal to $1/2$).

We propose to use the same protocol, but avoid the pedantic assumption,
using no-signalling twice (once in each direction) and  
invoking perfect correlation (the PCP, which we argued above 
follows from no-signalling).  The argument is as follows.
By no signalling in the $E$-to-$S$ direction (i.e., by 
envariance of probabilities), 
$p^S(0|\psi_1) = p^S(0|\psi_2)$.  By perfect correlation (the PCP, 
argued for above),  $p^S(0|\psi_2) = p^E(1|\psi_2)$.  By no-signalling
in the $S$-to-$E$ direction, $p^E(1|\psi_2) = p^E(1|\psi_3) \equiv
p^E(1|\psi_1)$.  By perfect correlation, $p^E(1|\psi_1) = 
p^S(1|\psi_1)$.  Thus we have established 
$p^S(0|\psi_1) = p^S(1|\psi_1)$.

Implicit in our remarks about the PCP is that, if desired, both of these 
arguments can be translated into arguments entirely about the
probabilities of the states $\ket{0}\ket{0}, \ket{1}\ket{1},
\ket{0}\ket{1},\ket{1}\ket{0}$ in the superpositions $\ket{\psi_i}$;
the second argument then becomes shorter since the PCP need not
be invoked. The above formulation, however, makes it clear that they
can also be formulated in a way that refers to probabilities of measurement
outcomes on Stan's and Emma's sides;  these arguments
are not necessarily tied to the relative state view.  

Essentially the same arguments establish that when there are $N$ equal
$|\alpha_k|$, each probability must be $p_k = 1/N$.  Then Zurek
establishes that when the real parts of $\alpha_k$ are square roots of
rationals $m_k/M$, say, the associated probabilities must be $m_k/M$.
The argument is a straightforward use of a local ancillary system to
turn the state into a superposition of states each of which has
the real parts of
its amplitude equal to $\sqrt{1/M}$, the above results implying 
that these must have equal
probability, and envariance of probabilities (in our terminology,
no signalling from the system that has interacted with the ancilla, to
the one that hasn't).

The passage to arbitrary amplitudes involves an assumption of
continuity of the probabilities in the amplitudes, which Zurek does
not explicitly justify.  How it might be justified probably depends on
the interpretation of quantum mechanics one adopts.  If the quantum
formalism is viewed as a handy mathematical way of systematizing or
compactly ``representing'' the probabilities of experimental results,
continuity of probabilities might be a desirable desideratum for
representations---if a representation is a structure with a natural
topology, but neighborhoods in that topology do not imply neigborhoods in
probabilities, that suggests that the topological aspect of the
representation is superfluous (but perhaps it could be needed to
represent dynamics, system combination, or some other aspect of things
beyond the ``statics'' of probabilities).  In the present context, 
however, such an ``operational'' argument would be backwards, since we
are trying to derive probabilities.  It is not clear to us why one
would rule out discontinous probability assignments even though they
may seem ``pathological.'' 
Wallace \cite{Wallace2002a} has given some
arguments based on robustness of probabilities to small perturbations,
and perhaps something can be made of these; but the ``smallness'' of a
perturbation, surely meant as nearness to the identity operator in
some standard metric, itself would seem to need an ``operational,''
perhaps probabilistic, interpretation.  The trickiest part of Gleason's
theorem is an argument---using noncontextuality---that
probabilities must be continuous.  One might be able to avoid assuming
continuity by adapting this argument to the present context, where
probabilities are only being assigned to product states of a composite
system, and some limited noncontextuality results are available from
no-signalling arguments as discussed in justifying the 
perfect correlation principle.

Perhaps the most important challenge to the whole approach is the point
that it only gives the probabilities for Schmidt states
of bipartite systems in pure states.  One can, and Zurek does,
argue that especially within
a relative-state approach this is enough:  that any measurement process
on a quantum system should be viewed as resulting in the eigenspaces of 
the measured observable being ``Schmidt eigenspaces'' for the decomposition of 
a bipartite pure state of the system and the rest of the world, with the 
Schmidt spaces of the rest of the world taken to constitute its 
``pointer basis,'' and we need only assign probability
to such spaces in order to interpret quantum mechanics.

If we can argue that in such situations, my pointer variables will
always be well correlated with some variables sufficiently 
spacelike separated from
them (which could, perhaps, be variables of the ``measured system''
itself) that causality prevents a signal from connecting them
with me during the time necessary to manipulate these variables, 
then no faster-than-light signalling, rather than just no signalling, can be
adduced as the basis of the argument for quantum probabilities.  
One might be tempted to argue that by a unitary transformation, we 
can always get my pointer variables so correlated, but to argue
that this does not affect probabilities is, it seems to me, a use
of the powerful EATMIMTEO, or some kindred invariance principle.  
However, using causality to buttress the 
appeal of no-signalling appears less crucial within the relative 
state interpretation.  There, with probabilities assigned to orthogonal
components of the overall state vector, we may note that regardless of
spacelike separation or the time it takes to do manipulations, we may
perform a unitary on the system that has been measured, and it would
be extraordinarily bizarre, perhaps inconsistent, for the probabilities
of the {\em already-recorded} results of the measurement to be changed
by such an operation---where by the fact that the measurement result
is already recorded, we mean that orthogonal subspaces of the system
have gotten correlated with (roughly) orthogonal ``pointer'' subspaces
of the rest of the world, each describing a more or less definite 
experience of the observer.  However, at heart this may just be another
guise for an invariance assumption, if a particularly reasonable (and
certainly empirically supported) one.  Perhaps this sort of argument motivated
Zurek to reduce his stress on causality and signalling.  In this
argument, it is less the fact that one can signal from S to E via 
a unitary on E, and more the fact that the unitary apparently changes
probabilities of an ``event''---admittedly, the subjective event of
an observer's conscious perspective turning out to be the one associated
with one pointer subspace rather than another---that has already happened,
that motivates envariance of probabilities.

In conclusion, we concur with Zurek that in a framework with a
distinguished bipartite factorization of Hilbert space, Zurek's notion
of envariance of a given bipartite state under a given local
transformation enables one to derive the quantum probabilities for the
Schmidt basis states of a pure bipartite state, from the assumption of
``envariance of probabilities,'' which we argue can be motivated
as a case of 
``no signalling from either system to the other by means of local
transformations.''  We suggest modifying the second step of 
Zurek's argument slightly to avoid a ``pedantic assumption'' 
which, we argue, is stronger than it seems.  
And we discuss the appeal
of the envariance assumption, and the sufficiency of its conclusion for
describing the use we actually make of quantum mechanics for probabilistic
prediction.  Both of these are strongest within a relative state view, 
but still have some appeal from other points of view.  Indeed, Zurek's
original argument, with its use of envariance in one direction, and 
a ``pedantic'' symmetry assumption in the other, might be thought to 
use envariance of probabilities
in the direction in which, at least on a relative-state
interpretation, it is most plausible, and the pedantic assumption which
allows him to use envariance only in one direction might be thought to 
be best justified within the relative-state interpretation as well.
Some of these assumptions seem to be related to, if perhaps weaker than,
assumptions made by Wallace and Deutsch in their derivations of 
probabilities within the relative state interpretation.  Interesting
directions for further research remain, including clarification of the
relationship between Zurek's assumptions and those used by Deutsch and 
Wallace, and clarification of the justification for assigning probabilities
at all, on the relative state interpretation.

\acknowledgments
The question of whether one could
do a Zurek-like derivation with only no-signalling, and not the pedantic 
assumption, arose during extensive email discussions with Jerry Finkelstein,
and the protocol that does it was suggested by him;  I am very grateful 
for his permission and encouragement to use it here, and for many other valuable
insights from our discussions.  I am also grateful to Wojciech Zurek for 
extensive discussions of his paper.  I thank the US DOE for financial support.

\vspace*{-2mm}

\begin{thebibliography}{14}
\expandafter\ifx\csname natexlab\endcsname\relax\def\natexlab#1{#1}\fi
\expandafter\ifx\csname bibnamefont\endcsname\relax
  \def\bibnamefont#1{#1}\fi
\expandafter\ifx\csname bibfnamefont\endcsname\relax
  \def\bibfnamefont#1{#1}\fi
\expandafter\ifx\csname citenamefont\endcsname\relax
  \def\citenamefont#1{#1}\fi
\expandafter\ifx\csname url\endcsname\relax
  \def\url#1{\texttt{#1}}\fi
\expandafter\ifx\csname urlprefix\endcsname\relax\def\urlprefix{URL }\fi
\providecommand{\bibinfo}[2]{#2}
\providecommand{\eprint}[2][]{\url{#2}}

\bibitem[{\citenamefont{Zurek}(2003)}]{Zurek2003a}
\bibinfo{author}{\bibfnamefont{W.~H.} \bibnamefont{Zurek}},
  \bibinfo{journal}{Phys. Rev. Lett.} \textbf{\bibinfo{volume}{90}},
  \bibinfo{pages}{120404} (\bibinfo{year}{2003}).

\bibitem[{\citenamefont{Everett}(1957{\natexlab{a}})}]{Everett57a}
\bibinfo{author}{\bibfnamefont{H.~D.} \bibnamefont{Everett}},
  \bibinfo{journal}{Reviews of Modern Physics} \textbf{\bibinfo{volume}{29}},
  \bibinfo{pages}{454} (\bibinfo{year}{1957}{\natexlab{a}}),
  \bibinfo{note}{reprinted in De Witt and Graham (1973) op. cit.}

\bibitem[{\citenamefont{Everett}(1957{\natexlab{b}})}]{Everett57b}
\bibinfo{author}{\bibfnamefont{H.~D.} \bibnamefont{Everett}}
  (\bibinfo{year}{1957}{\natexlab{b}}), \bibinfo{note}{{P}h.D. thesis,
  Princeton University. Reprinted in De Witt and Graham (1973) op. cit.}

\bibitem[{\citenamefont{Barnum}(1990)}]{Barnum90a}
\bibinfo{author}{\bibfnamefont{H.}~\bibnamefont{Barnum}}, \bibinfo{journal}{The
  many-worlds interpretation of quantum mechanics: psychological versus
  physical bases for the multiplicity of {"}worlds{"}}  (\bibinfo{year}{1990}),
  \bibinfo{note}{hardcopy available from the author on request.}

\bibitem[{\citenamefont{Caves et~al.}(2002)\citenamefont{Caves, Fuchs, and
  Schack}}]{Caves2002a}
\bibinfo{author}{\bibfnamefont{C.~M.} \bibnamefont{Caves}},
  \bibinfo{author}{\bibfnamefont{C.~A.} \bibnamefont{Fuchs}}, \bibnamefont{and}
  \bibinfo{author}{\bibfnamefont{R.}~\bibnamefont{Schack}},
  \bibinfo{journal}{Physical Review A} \textbf{\bibinfo{volume}{65}},
  \bibinfo{pages}{022305} (\bibinfo{year}{2002}).

\bibitem[{\citenamefont{Savage}(1972)}]{Savage72a}
\bibinfo{author}{\bibfnamefont{L.~J.} \bibnamefont{Savage}},
  \emph{\bibinfo{title}{The foundations of statistics}}
  (\bibinfo{publisher}{Dover}, \bibinfo{address}{New York},
  \bibinfo{year}{1972}), \bibinfo{edition}{2nd} ed.

\bibitem[{\citenamefont{Fishburn}(1970)}]{Fishburn70a}
\bibinfo{author}{\bibfnamefont{P.~C.} \bibnamefont{Fishburn}},
  \emph{\bibinfo{title}{Utitility theory for decision making}},
  vol.~\bibinfo{volume}{18} of \emph{\bibinfo{series}{Publications in
  Operations Research}} (\bibinfo{publisher}{John Wiley \& Sons},
  \bibinfo{address}{New York}, \bibinfo{year}{1970}).

\bibitem[{\citenamefont{Kreps}(1988)}]{Kreps88a}
\bibinfo{author}{\bibfnamefont{D.~M.} \bibnamefont{Kreps}},
  \emph{\bibinfo{title}{Notes on the theory of choice}}, Underground Classics
  in Economics (\bibinfo{publisher}{Westview Press}, \bibinfo{address}{Boulder
  and London}, \bibinfo{year}{1988}).

\bibitem[{\citenamefont{Pitowsky}(2003)}]{Pitowsky2003a}
\bibinfo{author}{\bibfnamefont{I.}~\bibnamefont{Pitowsky}},
  \bibinfo{journal}{Studies in History and Philosophy of Modern Physics}
  \textbf{\bibinfo{volume}{34}}, \bibinfo{pages}{295} (\bibinfo{year}{2003}).

\bibitem[{\citenamefont{Deutsch}(1999)}]{Deutsch99a}
\bibinfo{author}{\bibfnamefont{D.}~\bibnamefont{Deutsch}},
  \bibinfo{journal}{Proc R Soc London A} \textbf{\bibinfo{volume}{455}},
  \bibinfo{pages}{93129} (\bibinfo{year}{1999}).

\bibitem[{\citenamefont{Wallace}(2002)}]{Wallace2002a}
\bibinfo{author}{\bibfnamefont{D.}~\bibnamefont{Wallace}},
  \bibinfo{journal}{arXiv.org e-print quant-ph/0211014}
  (\bibinfo{year}{2002}).

\bibitem[{\citenamefont{Wallace}(2003)}]{Wallace2003a}
\bibinfo{author}{\bibfnamefont{D.}~\bibnamefont{Wallace}},
  \bibinfo{journal}{Studies in History and Philosophy of Modern Physics}
  \textbf{\bibinfo{volume}{34}}, \bibinfo{pages}{415} (\bibinfo{year}{2003}).

\bibitem[{\citenamefont{Saunders}(2002)}]{Saunders2002a}
\bibinfo{author}{\bibfnamefont{S.}~\bibnamefont{Saunders}},
  \bibinfo{journal}{arXiv.org e-print quant-ph/0212238}
  (\bibinfo{year}{2002}).

\bibitem[{\citenamefont{Barnum et~al.}(2000)\citenamefont{Barnum, Caves,
  Finkelstein, Fuchs, and Schack}}]{Barnum2000a}
\bibinfo{author}{\bibfnamefont{H.}~\bibnamefont{Barnum}},
  \bibinfo{author}{\bibfnamefont{C.~M.} \bibnamefont{Caves}},
  \bibinfo{author}{\bibfnamefont{J.}~\bibnamefont{Finkelstein}},
  \bibinfo{author}{\bibfnamefont{C.~A.} \bibnamefont{Fuchs}}, \bibnamefont{and}
  \bibinfo{author}{\bibfnamefont{R.}~\bibnamefont{Schack}},
  \bibinfo{journal}{Proceedings of the Royal Society of London A}
  \textbf{\bibinfo{volume}{456}}, \bibinfo{pages}{1175} (\bibinfo{year}{2000}).

\end{thebibliography}

\end{document}